\documentclass[a4paper]{article}
\usepackage{epsfig,amsmath,amsfonts,amssymb,setspace,multirow,textcomp}
\usepackage{float}
\usepackage[T1]{fontenc}
\makeatletter
\renewcommand{\@oddhead}{\textit{Advances in Astronomy and Space Physics} \hfil}
\renewcommand{\@evenfoot}{\hfil \thepage \hfil}
\renewcommand{\@oddfoot}{\hfil \thepage \hfil}
\makeatother

\renewenvironment{thebibliography}[1]{\begin{oldthebibliography}{#1}\setlength{\parskip}{0ex}\setlength{\itemsep}{0ex}}{\end{oldthebibliography}}
\usepackage[utf8]{inputenc}
\usepackage[english,ukrainian]{babel}

\begin{document}
\fontsize{11}{11}\selectfont
\title{Third components with elliptical orbits in the eclipsing binaries: AB Cas, AF Gem, AR Boo, BF Vir and CL Aur}
\author{\textsl{D.E. Tvardovskyi$^{1, 2}$}}
\date{\vspace*{-6ex}}
\maketitle

\begin{center} {\small $^{1}$Odessa I. I. Mechnikov National University,  Dvoryanskaya str., 2, 65082, Odessa, Ukraine\\* $^2$ Department “Mathematics, Physics and Astronomy”, Odessa National Maritime University, Mechnikova str., 34, 65029 , Odessa, Ukraine\\*}
{\tt dmytro.tvardovskyi@gmail.com}
\end{center}

\begin{abstract}
In this research, five eclipsing binary stars were studied: AB Cas, AF Gem, AR Boo, BF Vir and CL Aur. The large sets of moments of minima were used: from the international BRNO database and amateur observations from the database AAVSO. Firstly, moments of minima for AAVSO observations were obtained (totally – 222 minima). The software MAVKA was used. It was kindly provided by Kateryna D. Andrych and Ivan L. Andronov (2019OEJV..197...65A) and approximation with various methods in order to find the best fit. Then all obtained moments of minima were combined and O-C diagrams were plotted. For all stars these diagrams represented sinusoidal-like oscillations with superposition of parabolic trend. One of the possible reasons for such oscillations could be presence of well-known light-time effect (LTE) caused by third component with elliptical orbit. Parabolic trend was explained as mass transfer between components of binary system. For all these stars we computed possible mass of the third component, orbital elements, mass transfer rate and errors for all computed values. 
\end{abstract}
\section*{\sc Introduction}
For this research five eclipsing binaries were chosen: AB Cas, AF Gem, AR Boo, BF Vir and CL Aur. All of them are well-known stellar systems and were observed during long period of time [12]. To start the research, period and initial epoch were taken from General Catalogue of Variable Stars (GCVS [21]). Masses of the binary systems were taken from previously published articles of different authors.

\begin{table}[H]
 \caption{The characteristics of some identified stars.}\label{tab1}
 \vspace*{1ex}
\begin{center}
\begin{tabular}{p{1.5cm}p{2.5cm}p{2cm}p{2cm}p{2cm}p{1.5cm}}
\hline
Stellar system & Initial epoch (JD-2400000) & Period (days) & $M_{1}, M_{\odot}$ & $M_{2}, M_{\odot}$ & Reference\\
\hline
AB Cas & 56541.878 & 1.366892 & 2.01 $\pm$ 0.02 & 0.37 $\pm$ 0.02 & [15]\\
AF Gem & 27162.3095 & 1.24350348 & 2.92 $\pm$ 0.05 & 1.04 $\pm$ 0.02 & [31]\\
AR Boo & 52500.3838 & 0.34487642 & 0.90 $\pm$ 0.06 & 0.35 $\pm$ 0.02 & [17]\\
BF Vir & 46070.684 & 0.64057 & 2.04 $\pm$ 0.14 & 1.02 $\pm$ 0.07 & [19]\\
CL Aur & 50097.2712 & 1.24437505 & 2.24 $\pm$ 0.16 & 1.35 $\pm$ 0.09 & [18]\\
\hline
\end{tabular}
\end{center}
\end{table}

For AR Boo, BF Vir and CL Aur errors of binary system masses were not provided by any author. Thus, they were considered as 7\% of each individual star mass.
\section*{\sc O-C diagrams and bibliography analysis}
Note: the $\pm1\sigma$ and $\pm2\sigma$ confidence intervals are shown, where $\sigma$ is an unbiased estimate or the r.m.s. deviation of the points from the fit.
\begin{center}
{\bf AB Cas}
\end{center}

\begin{figure}[H]
\centering
\includegraphics[width=75mm]{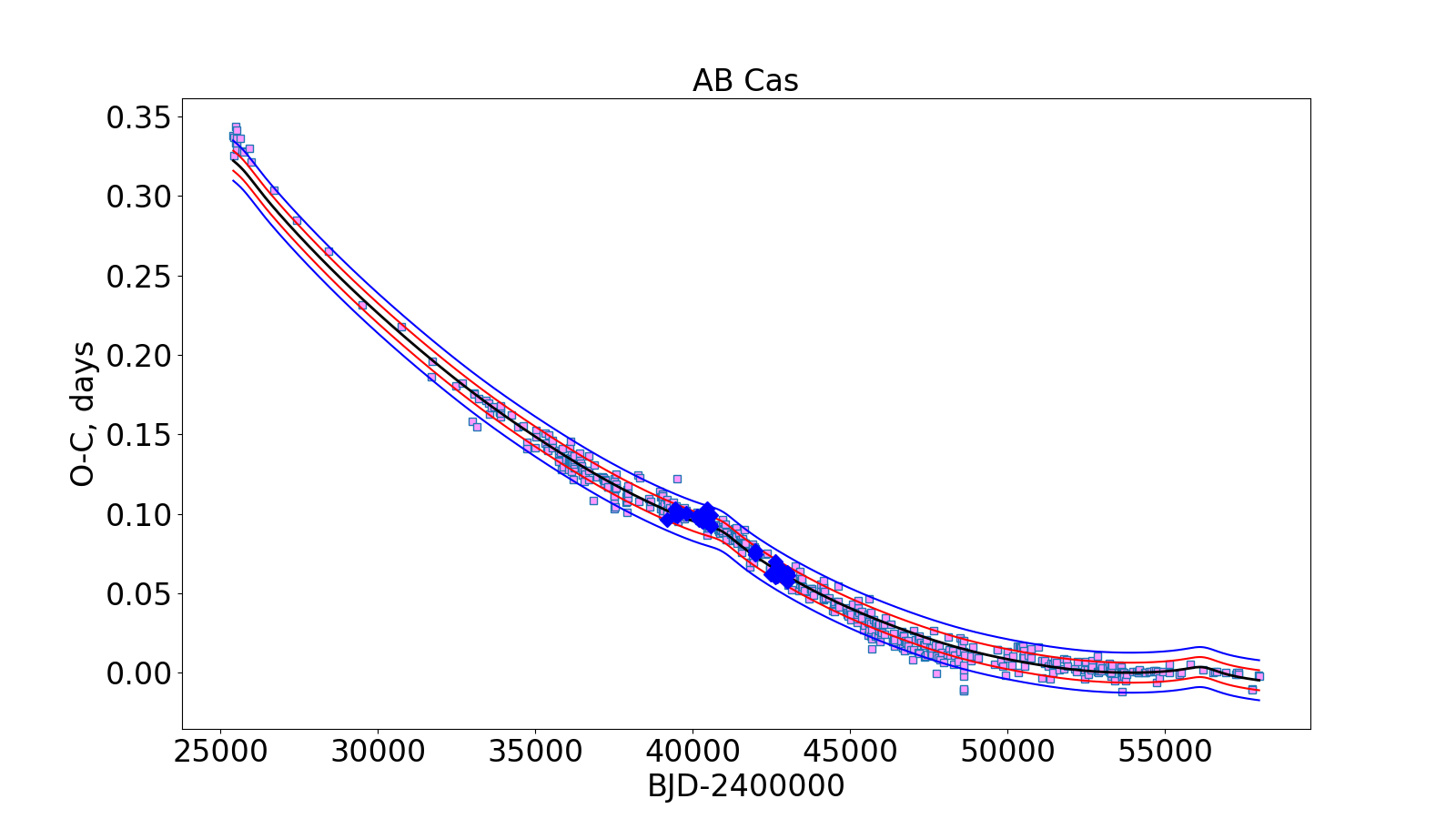}
\caption{O-C curve of AB Cas}
\end{figure}
Mass transfer was supposed in [23], [16] and [24] and its rate was calculated there. Third component was supposed and its mass computed in [23], [16], [2] and [24]. Orbital elements of the third components orbit were computed in [2] and [24].

\begin{center}
{\bf AF Gem}
\end{center}

\begin{figure}[H]
\centering
\includegraphics[width=75mm]{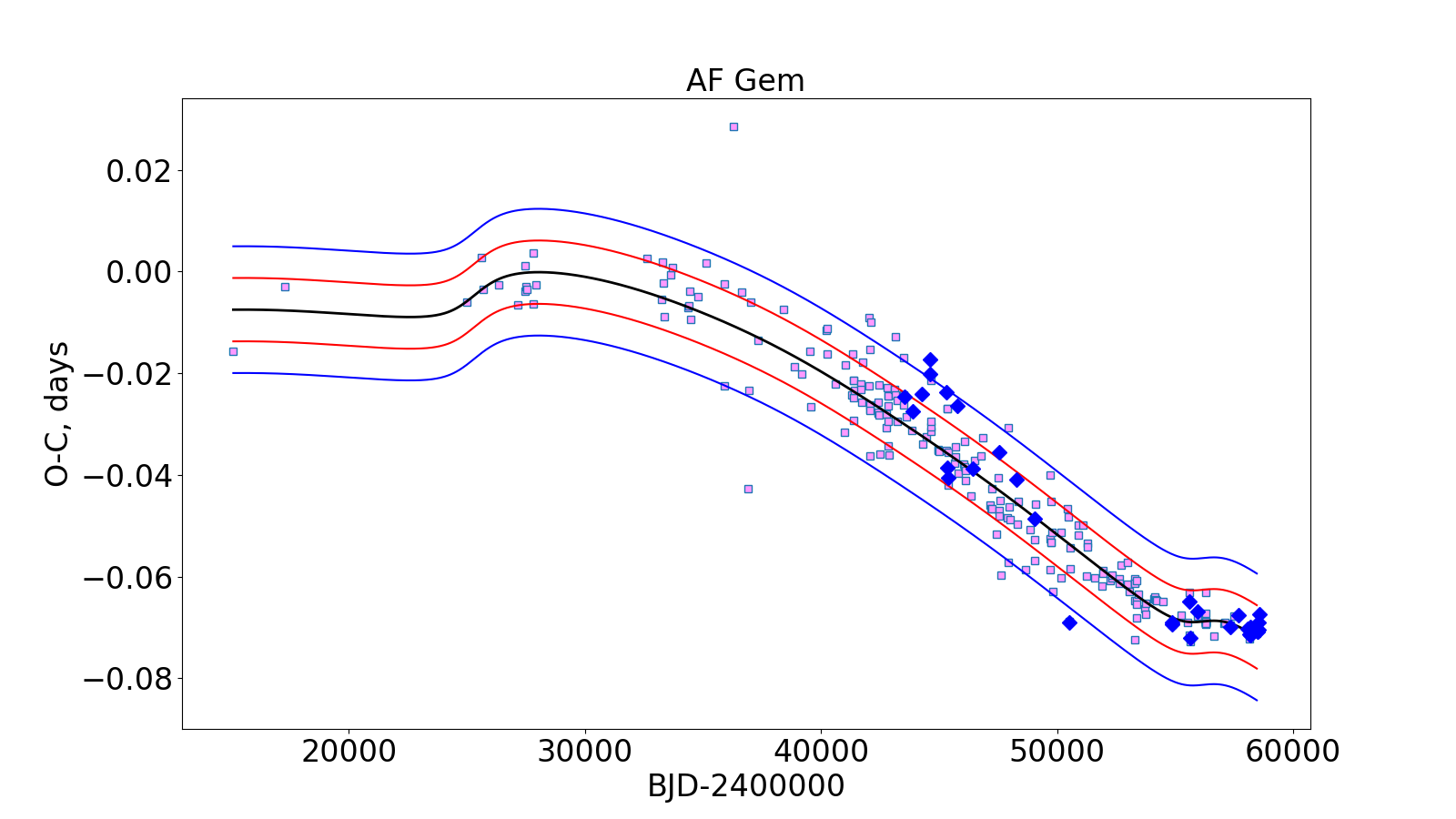}
\caption{O-C curve of AF Gem}
\end{figure}
Third and forth components were supposed in [31] and their masses were estimated.

\begin{center}
{\bf AR Boo}
\end{center}

\begin{figure}[H]
\centering
\includegraphics[width=75mm]{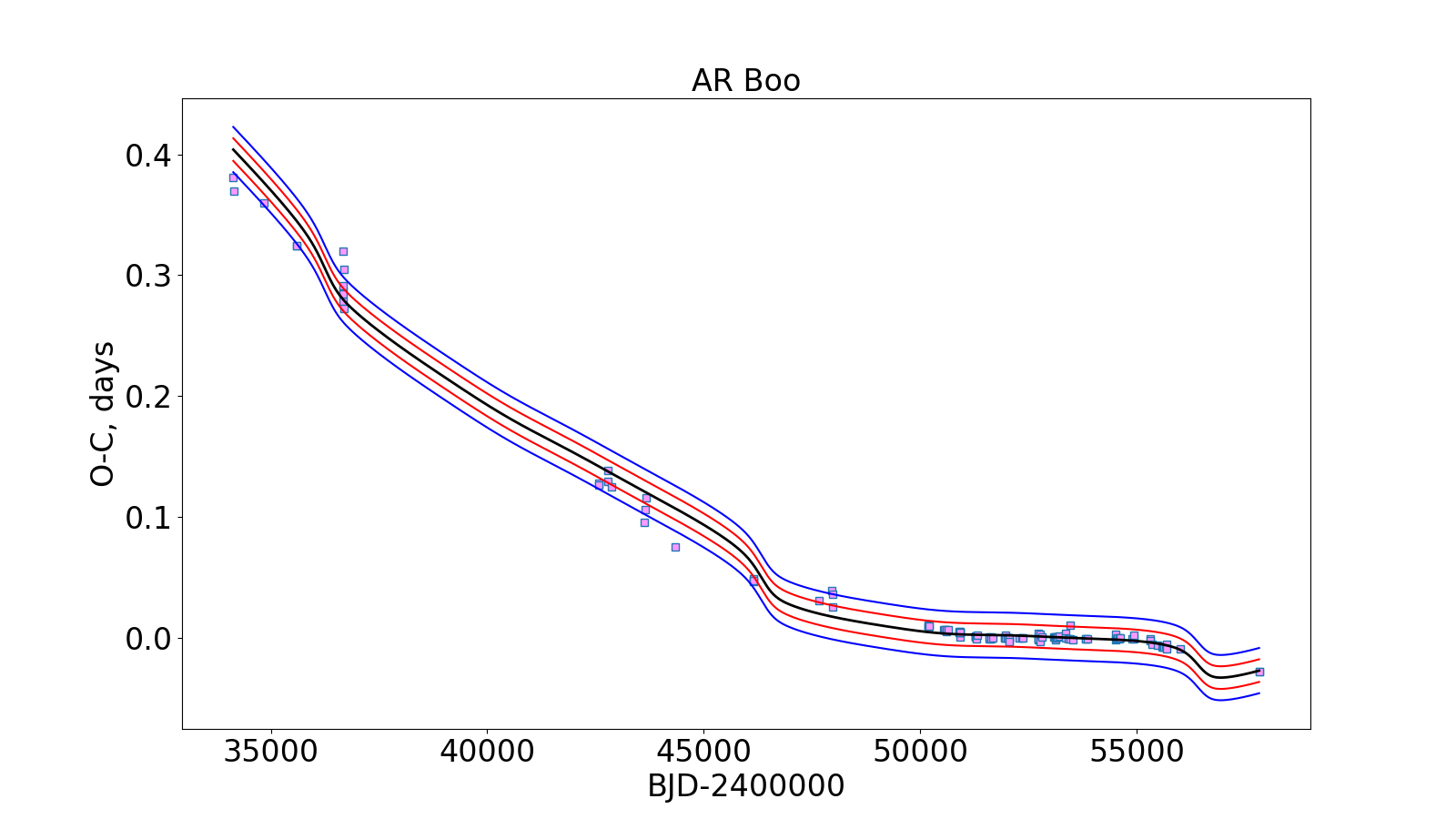}
\caption{O-C curve of AR Boo}
\end{figure}
Mass transfer was supposed in [17]. Third component was supposed in [20] and [22]. Third components mass was calculated only in [17]. Orbital elements were not estimated in any article.

\begin{center}
{\bf BF Vir}
\end{center}

\begin{figure}[H]
\centering
\includegraphics[width=75mm]{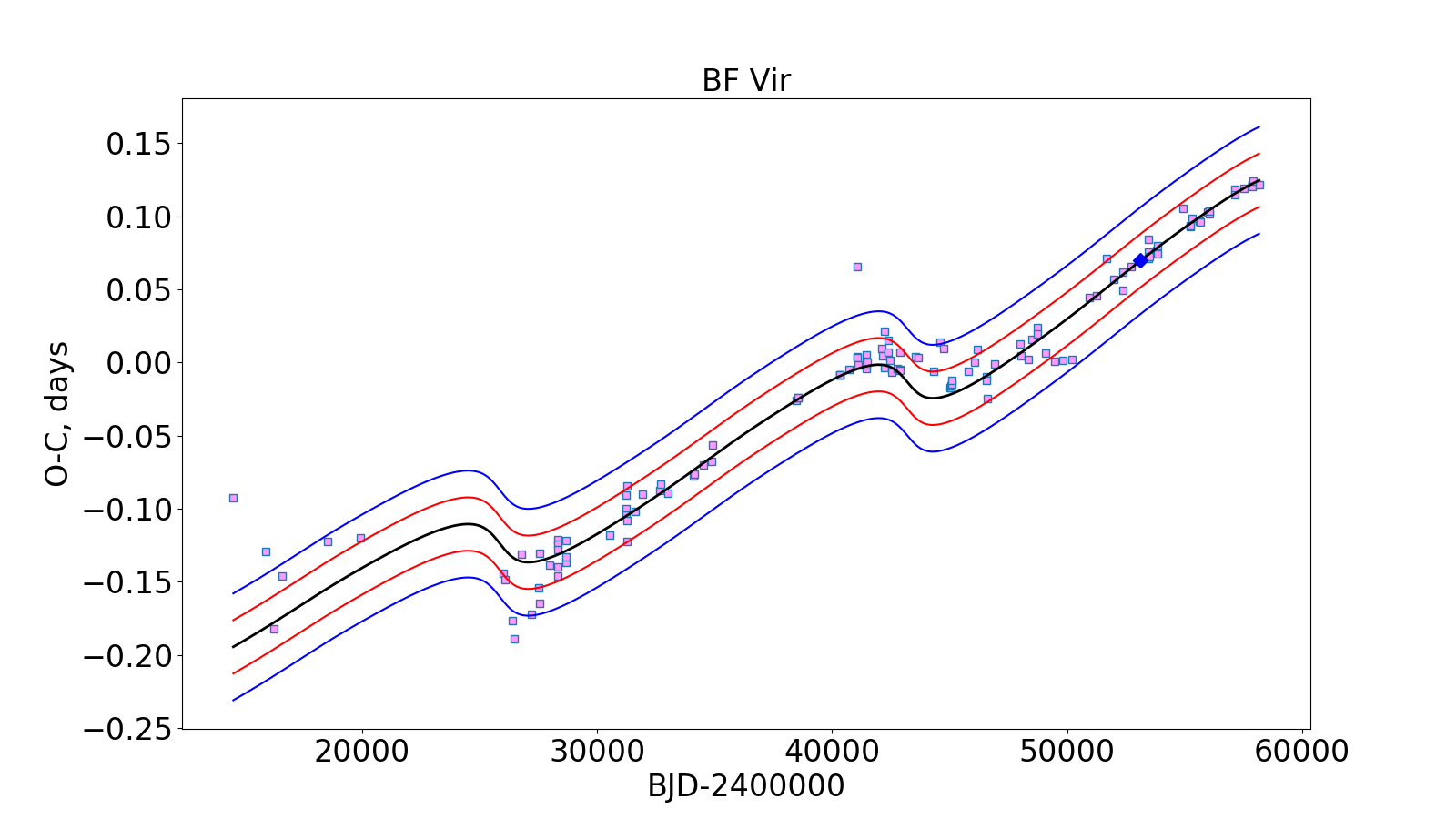}
\caption{O-C curve of BF Vir}
\end{figure}
Mass transfer rate was calculated in [19]. In addition, in this article mass and orbital elements of the third component were obtained. In another article, [22] only third component mass was obtained.

\begin{center}
{\bf CL Aur}
\end{center}

\begin{figure}[H]
\centering
\includegraphics[width=75mm]{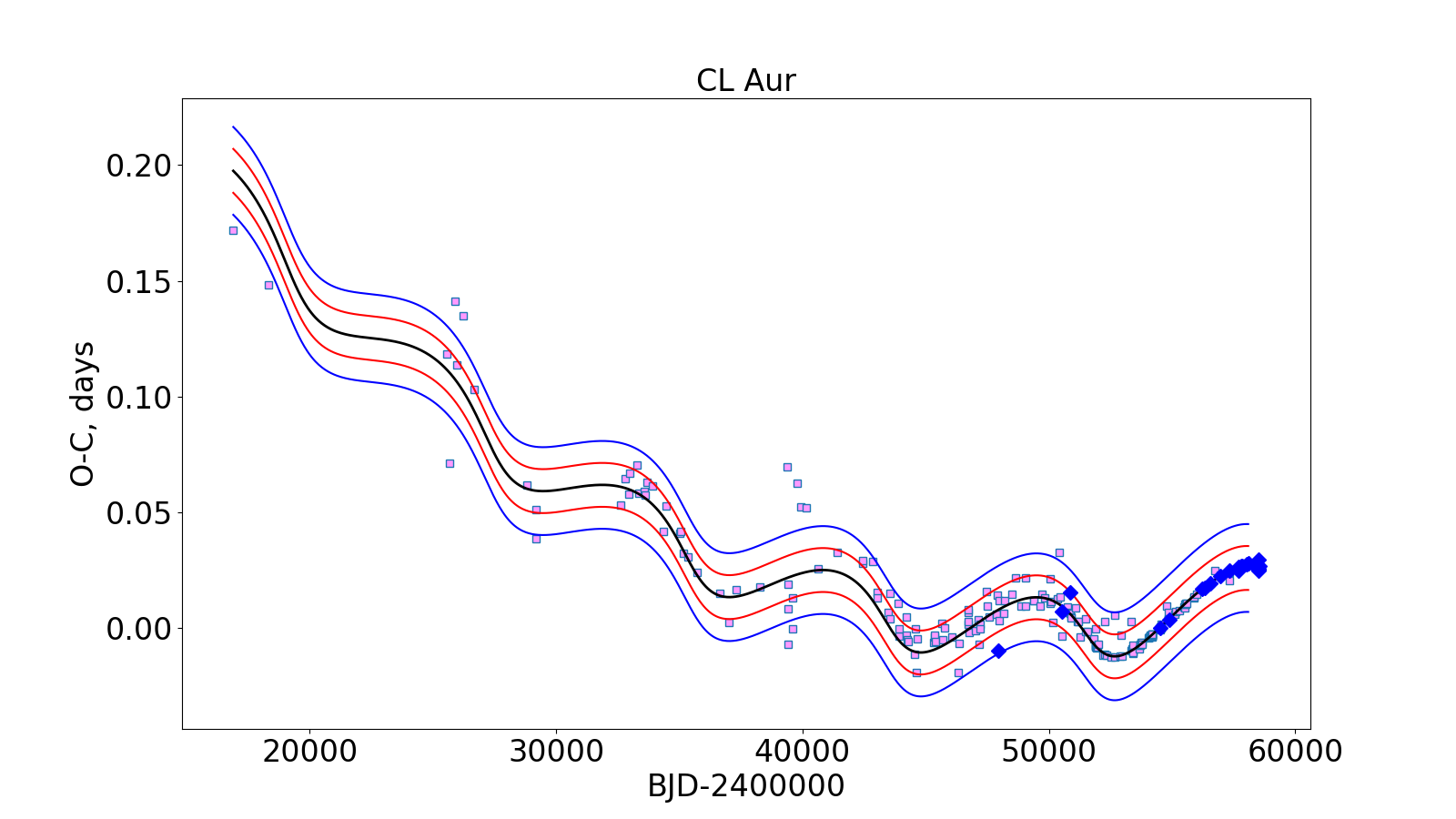}
\caption{O-C curve of CL Aur}
\end{figure}
Mass transfer rate was not calculated in any article before. In [30] and [31] both mass of the third component was calculated. In [18] just third component's mass was estimated.
\section*{\sc Methods and algorithms}
To compute the parameters of the third component and its orbit (as well as mass transfer rate) the classical O-C analysis was used. From parameters of parabolic trend we computed rate of the mass transfer. From oscillations orbital elements and third components mass were obtained. Both trend and oscillations were simulated simultaneously.\\
Details about algorithm O-C analysis and formulas for calculation were described in our previous articles: [25] and [26]. Moments of minima which were obtained using AAVSO observations were really helpful for investigation. Total number of data points increased on 15\%.

\section*{\sc Analysis of MAVKA processing results}
Software MAVKA was widely used in this research. Previous investigation and development of MAVKA algorithms was done in [3], [6], [7], [11], [13]. In the final version of this software there are 9 methods of approximation. Firstly each minimum was processed with all nine ones. Then, for each minimum the approximation with the best accuracy was chosen. Finally, the table with all moments of minima taken from AAVSO database was created.
Space limitation does not allow us to provide the whole table with all extrema. However, the general description of the results is provided:

\begin{table}[H]
\caption{General description of amount of obtained minima}
\vspace*{1ex}
\begin{center}
\begin{tabular}{p{2.5cm}p{3cm}p{3cm}}
\hline
Stellar system & BRNO minima & This article minima \\
\hline
AB Cas & 763 & 169\\
AF Gem & 247 & 30\\
AR Boo & 118 & 0\\
BF Vir & 134 & 1\\
CL Aur & 228 & 22\\
\hline
\end{tabular}
\end{center}
\end{table}

\section*{\sc Discussion and conclusions}

\begin{table}[H]
\caption{Orbital elements, masses of the third components and mass transfer rates for all 5 stars}
\vspace*{1ex}
\begin{center}
\begin{tabular}{p{2.5cm}p{2.2cm}p{2.2cm}p{2.2cm}p{2.2cm}p{2.2cm}}
\hline
Value & AB Cas & AF Gem & AR Boo & BF Vir & CL Aur \\
\hline
$\alpha$, $10^{-12} \space {\rm days^{-1}}$ & 306.7 $\pm$ 2.4 & 53.7 $\pm$ 5.7 & 903.1 $\pm$ 22.3 & 37.8 $\pm$ 6.4 & 175.6 $\pm$ 4.5 \\
$\beta$, $10^{-6}$ & 35.4 $\pm$ 0.2 & 2.3 $\pm$ 0.5 & 99.9 $\pm$ 2.2 & 3.8 $\pm$ 0.6 & 17.3 $\pm$ 0.4 \\
$\gamma$, days & 1.023 $\pm$ 0.005 & 0.029 $\pm$ 0.009 & 2.740 $\pm$ 0.049 & 0.274 $\pm$ 0.015 & 0.424 $\pm$ 0.010 \\
$a \sin{i}, {\rm \space 10^6 km}$ & 176 $\pm$ 7 & 232 $\pm$ 18 & 742 $\pm$ 102 & 847 $\pm$ 174 & 374 $\pm$ 6 \\
e, 1 & 0.848 $\pm$ 0.044 & 0.758 $\pm$ 0.038 & 0.843 $\pm$ 0.038 & 0.714 $\pm$ 0.117 & 0.333 $\pm$ 0.023 \\
$\omega$, rad & 4.96 $\pm$ 0.07 & 6.13 $\pm$ 0.13 & 3.41 $\pm$ 0.05 & 0.06 $\pm$ 0.09 & 0.43 $\pm$ 0.08 \\
$t_{0}$, JD-2400000 & 4826 $\pm$ 332 & 51999 $\pm$ 3347 & 4010 $\pm$ 832 & 8757 $\pm$ 790 & 33449 $\pm$ 700 \\
T, days & 15264 $\pm$ 84 & 30877 $\pm$ 934 & 10082 $\pm$ 142 & 17330 $\pm$ 317 & 8111 $\pm$ 70 \\
$\dot{M}$, $10^{-9} \space \frac{M_{\odot}}{\rm year}$ & 24.8 $\pm$ 1.7 & 17.0 $\pm$ 1.9 & 365 $\pm$ 46 & 29.3 $\pm$ 6.7 & 117 $\pm$ 24.2 \\
$M_{3}$, $M_{\odot}$ & 0.19 $\pm$ 0.01 & 0.21 $\pm$ 0.02 & 0.91 $\pm$ 0.19 & 1.12 $\pm$ 0.30 & 0.86 $\pm$ 0.07 \\
\hline
\end{tabular}
\end{center}
\end{table}

\begin{figure}[H]
\centering
\includegraphics[width=120mm]{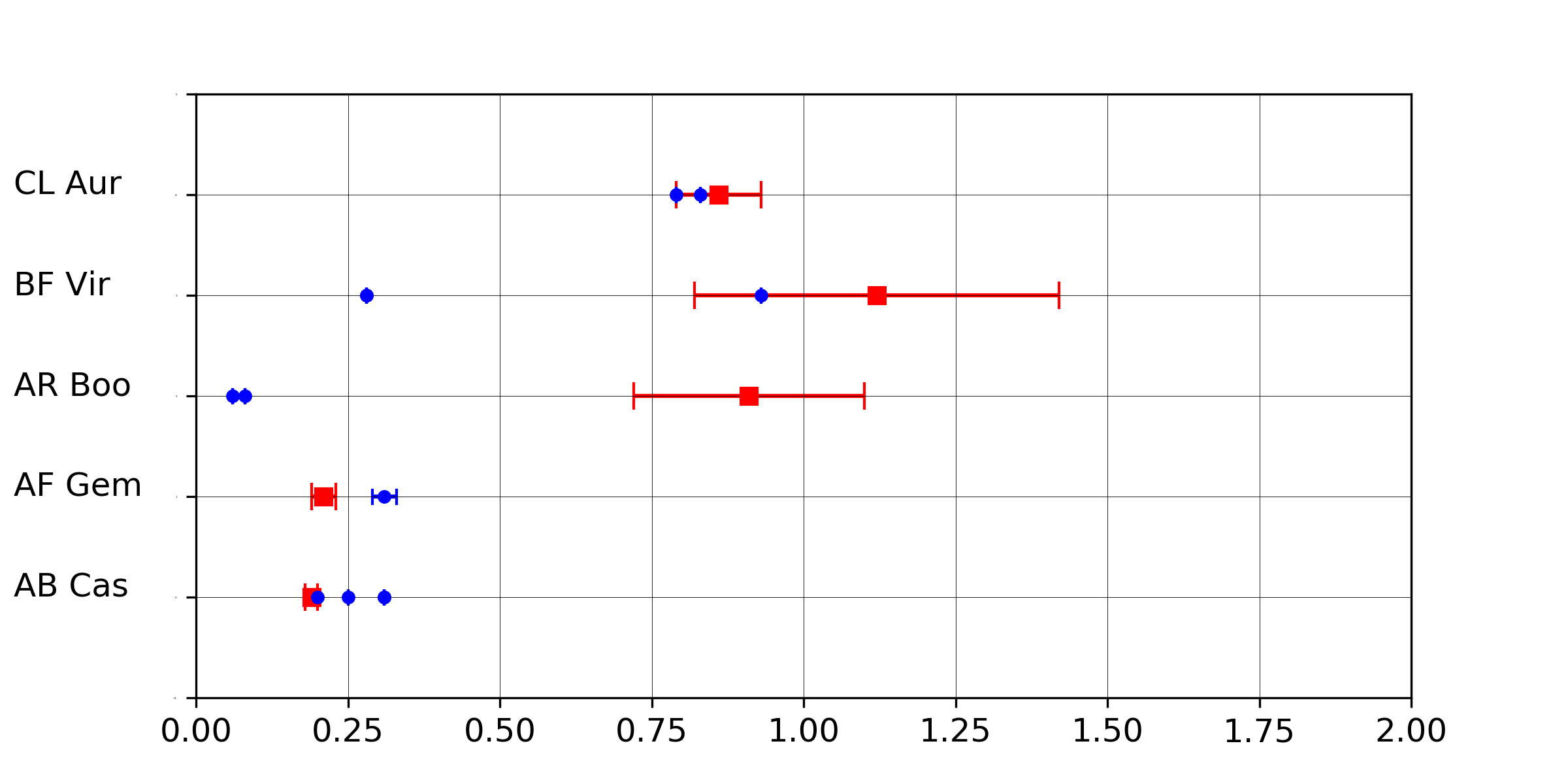}
\caption{Computed masses of the third components (squares) in comparison with previous results (dots)}
\end{figure}
Results in this article generally are not in good agreement with results of other authors. Generally disagreement is several times larger than error bars. Such large deviation could be explained due to different orbital models and larger data set we used for current calculations. All of third components (according to their masses) are low-mass stars with highly-eccentric orbits (average eccentricity is almost 0.7). \\
Mass transfer rates also are not in good agreement with previous researches. For AB Cas it is about 35-50 \% of the values, published in [23], [16] and [24]. For AR Boo the situation is opposite: computed mass transfer rate is twice larger than in [17]. So large deviation could be due to different ephemerides used in various researches. \\ 
Orbital elements for AB Cas is quite different from published in [2], [24]. Average deviation is about 25\%. For BF Vir obtained values are almost the same that ones in [19] and some of them even are within errors of calculation. For CL Aur computed orbital elements mostly are within error bars ([29], [30]), except $\omega$ that differs almost for $\pi$/2 rad.

\section*{\sc Acknowledgments}
\indent \indent We sincerely thank the AAVSO (Kavka, 2018) [1] and BRNO [7] associations of variable stars observers for their work that has made this research possible. We are grateful to I.L.Andronov for fruitful discussons. In addition, we are grateful to We are also grateful to K.D. Andrych, I.L.Andronov and their coauthors [8], [9], [10] who provided software MAVKA for obtaining moments of minima.\\
This research was done as the part of the projects Inter-Longitude Astronomy [4], [5], UkrVO [27] and AstroInformatics [28] as well as previous research.


\end{document}